\documentclass[12pt]{iopart}
\pdfoutput=1
\usepackage{amssymb,amstext,graphicx}
\usepackage[numbers,square,sort,compress]{natbib}
\usepackage{xcolor,soul}

\newcommand{\al}{\alpha}

\newcommand{\vhi}{\varphi}

\newcommand{\sig}{\sigma}

\newcommand{\Nb}{N_\text{b}}
\newcommand{\Nc}{N_\text{c}}
\newcommand{\x}{\mathbf r}
\newcommand{\ta}{\tau_\alpha}
\newcommand{\tx}{\tau_\text{x}}
\newcommand{\tf}{\tau_\text{f}}
\newcommand{\tobs}{t_\text{obs}}
\newcommand{\Dl}{D_\text{L}}
\newcommand{\pb}{P_\text{B}}
\newcommand{\pbm}{P^\text{m}_\text{B}}

\begin{document}

\title[Crystallization of polydisperse hard spheres]{Polydisperse hard spheres: Crystallization kinetics in small systems and role of local structure}

\author{Matteo Campo and Thomas Speck}
\address{Graduate School Materials Science in Mainz, Staudinger Weg 9, 55128
  Mainz, Germany}
\address{Institut f\"ur Physik, Johannes Gutenberg-Universit\"at Mainz,
  Staudingerweg 7-9, 55128 Mainz, Germany}

\begin{abstract}
  We study numerically the crystallization of a hard-sphere mixture with 8\% polydispersity. Although often used as a model glass former, for small system sizes we observe crystallization in molecular dynamics simulations. This opens the possibility to study the competition between crystallization and structural relaxation of the melt, which typically is out of reach due to the disparate timescales. We quantify the dependence of relaxation and crystallization times on density and system size. For one density and system size we perform a detailed committor analysis to investigate the suitability of local structures as order parameters to describe the crystallization process. We find that local structures are strongly correlated with generic bond order and add little information to the reaction coordinate.
\end{abstract}


\section{Introduction}

\begin{figure}[t]
  \centering
  \includegraphics{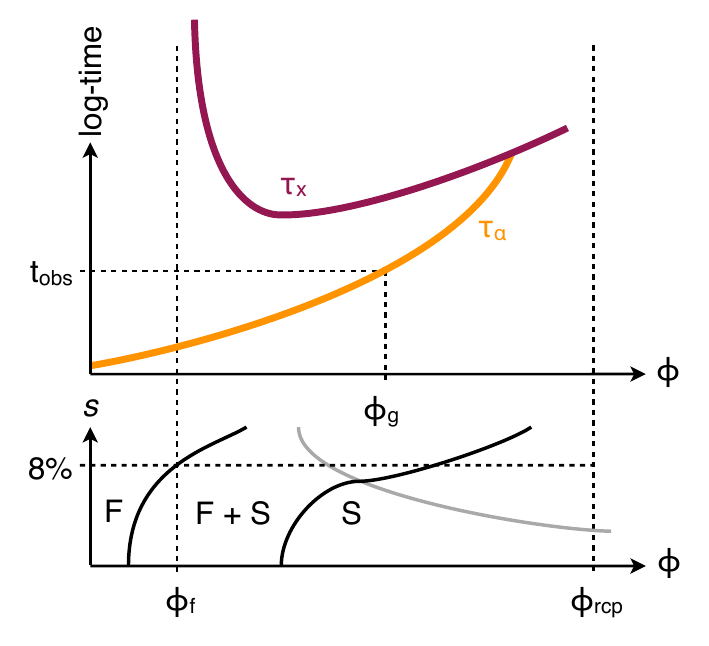}
  \caption{Sketch of timescales (top, adapted from Ref.~\citenum{chan10}) and the equilibrium phase diagram for hard spheres (bottom). There is a first-order phase transition from fluid (F) to solid (S) with both phases coexisting (F+S) in between. Polydispersity as measured by $s$ shifts the phase boundaries and at larger polydispersity fractionation occurs (gray line)~\cite{faso03}. The crystallization time $\tx$ is non-monotonous. It diverges at the freezing density $\phi_\text{f}$ before reaching a minimum and again increases due to slow diffusion. The structural relaxation time $\ta$ monotonously increases and might intersect the crystallization time (``kinetic spinodal''). For $\phi>\phi_\text{g}$ a glass can form.}
  \label{fig:sketch}
\end{figure}

A comprehensive understanding of the glass transition is still an open issue~\cite{debe01,biro13}. Glasses are typically prepared by quenching a ``melt'', either by cooling a liquid or compressing a colloidal suspension. Passing through the fluid-solid transition, glass formation competes with crystallization (exception are, \emph{e.g.}, patchy particles~\cite{smal13} and idealized kinetically constrained models~\cite{rito03,chan10}). Glass formation is a dynamical process involving at least three timescales: the crystallization time $\tx$, the structural relaxation time $\ta$ of the melt, and the experimentally accessible time $\tobs$, see Fig.~\ref{fig:sketch} for a sketch. A necessary condition for glass formation thus is $\tobs<\ta\ll\tx$, with the precise condition at which the melt falls out of equilibrium depending on the quench protocol.

Hard spheres is one of the most studied model systems for the glass transition. In particular for monodisperse hard spheres crystallization has been investigated extensively, both in experiments on colloids~\cite{puse86,taff13,palb14} and computer simulations~\cite{auer01,kawa10,schi10a,fili10}. Mixtures of hard spheres with different sizes are routinely employed to avoid crystallization and to study the kinetic arrest~\cite{bram09,zacc15}. Moderate polydispersity $s\lesssim0.06$ (measured as the standard variation $s$ of particle sizes divided by the mean) seems to have little influence on phase behavior (phase boundaries are shifted to higher volume fraction) and single particle dynamics~\cite{zacc09,puse09}. Above $s\approx0.06$ fractionation, \emph{i.e.}, coexisting phases with different size distributions, has been predicted~\cite{faso03}. In Ref.~\citenum{zacc09} crystallization up to $s\leqslant0.07$ has been observed, and in Ref.~\citenum{puse09} it has been shown that the nucleation rates for small polydispersity collapse when normalized by the diffusion coefficient and plotted versus the supersaturation $\phi-\phi_\text{f}$.

Already in the 1950s Sir Charles Frank speculated that particles in the liquid pack locally into clusters (locally favored structures), contributing to the glass forming abilities since the rearrangement necessary to transform local structures into the crystal ``is quite costly of energy in small localities, and only becomes economical when extended over a considerable volume''~\cite{fran52}. Different locally favored structures, most notable the icosahedron, have been identified and shown to occur more frequently in the metastable fluid melt~\cite{cosl07,cosl11,mali13a}, see Ref.~\citenum{roya15} for a comprehensive review. However, to which degree there is indeed a causal link between local structures and (local) particle dynamics is debated~\cite{hock14,cosl16}.

Although the crystallization kinetics and mechanism of several model glass formers has been studied~\cite{toxv09,pede11}, including monodisperse hard spheres~\cite{sanz11,sanz14} and colloidal suspensions~\cite{gold16}, a conclusive picture remains elusive. Here we analyse finite-size effects, which is an important tool of computational statistical mechanics. To this end we study a polydisperse hard-sphere mixture with $s\simeq0.08$ with system sizes for which crystallization times move into the range accessible by straightforward computer simulations.

\section{Model and methods}

We study a five-component equimolar mixture of $N$ hard spheres in a cubic box of volume $V$ with periodic boundary conditions. The mixture is the same that has been studied in Ref.~\citenum{dunl15}. Particles have equal masses $m$ and diameters $(0.799\sig,0.861\sig,0.899\sig,0.938\sig,\sig)$ with $s\simeq0.08$. We employ event-driven molecular dynamics simulations at constant volume and report all times in units of $\sqrt{m\sig^2/k_\text{B}T}$ and lengths in units of $\sig$. The equilibrium phase diagram is determined by the packing fraction $\phi=V_\text{hs}/V$ and the polydispersity $s$, see Fig.~\ref{fig:sketch}, where $V_\text{hs}=\frac{\pi}{6}\sum_{k=1}^N\sig_k^3$ is the volume occupied by the hard spheres.

\subsection{Structural relaxation of the supersaturated fluid}

\begin{figure}[t]
  \centering
  \includegraphics{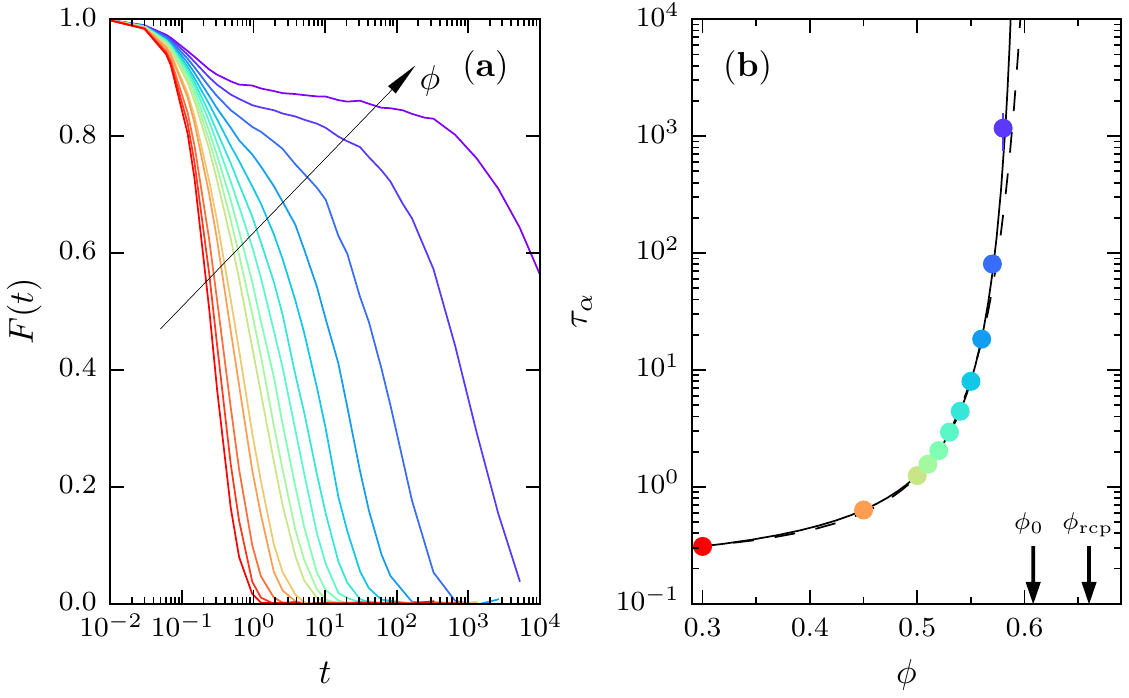}
  \caption{(a)~Intermediate scattering functions $F(t)$ [Eq.~(\ref{eq:isf})] for several volume fractions $0.30\leqslant\phi\leqslant 0.59$ and $N=1000$ hard spheres. (b)~Extracted structural relaxation times $\ta$ as a function of volume fraction. The solid line is a fit of Eq.~(\ref{eq:tau}) with exponent $\delta=1$, which diverges at $\phi_0\simeq0.61$. The dashed line is an alternative fit with $\phi_0=0.66$ corresponding to random close packing. Colors are the same as in (a).}
  \label{fig:glass}
\end{figure}

To characterize the dynamical behavior of the fluid we first study a system with $N=1000$ particles. We describe the dynamics through the self-intermediate scattering function
\begin{equation}
  \label{eq:isf}
  F(t) = \left\langle \frac{1}{N}\sum_{k=1}^N e^{i\mathbf k\cdot[\x_k(t)-\x_k(0)]} \right\rangle
\end{equation}
calculated from long trajectories at wave vector $|\mathbf k|=2\pi$ close to the first peak of the static structure factor, see Fig.~\ref{fig:glass}(a). Here, $\x_k(t)$ is the position of particle $k$ at time $t$. Using the bond-order parameter introduced in the next section, we confirm the absence of crystallinity for this system size for all packing fractions studied. The structural relaxation times $\ta$ are extracted from $F(\ta)=1/e$ through interpolation and are plotted in Fig.~\ref{fig:glass}(b).

The range over which hard spheres form random packings is obviously limited. Random close packing $\phi_\text{rcp}$ defines the densest amorphous packings possible, with a sharp onset of local crystallinity~\cite{kapf12} and a diverging pressure of the (metastable) fluid branch (``jamming'')~\cite{biro07}. For monodisperse hard spheres $\phi_\text{rcp}\simeq0.64$, while for $s\simeq0.08$ it has been estimated to be larger, $\phi_\text{rcp}\simeq0.66$~\cite{scha94}. Relaxation times are typically fitted to an expression of the form
\begin{equation}
  \label{eq:tau}
  \ta(\phi) = \tau_0\exp\left[\frac{A}{(\phi_0-\phi)^\delta}\right]
\end{equation}
with kinetic arrest occurring at packing fraction $\phi_0$, which finds support from several theories with varying predictions for the exponent $\delta$. A fundamental question is whether the kinetic arrest coincides with random close packing, $\phi_0=\phi_\text{rcp}$ (as expected from free volume arguments~\cite{kami07} and within dynamical facilitation~\cite{isob16}), or whether $\phi_0<\phi_\text{rcp}$ at a pressure that is still finite~\cite{pari10}. In our case a free fit yields an exponent close to unity, so first we fix $\delta=1$ for which we obtain $\phi_0\simeq0.61<\phi_\text{rcp}$ in agreement with Ref.~\citenum{dunl15}. We then fix $\phi_0=0.66\simeq\phi_\text{rcp}$ and again fit the data. We now obtain $\delta\simeq2.0$, which agrees with Ref.~\citenum{bram09}. Both fits are very close and only differ appreciably for the highest packing fraction, for which the error is also the largest.

\begin{figure}[t]
  \centering
  \includegraphics{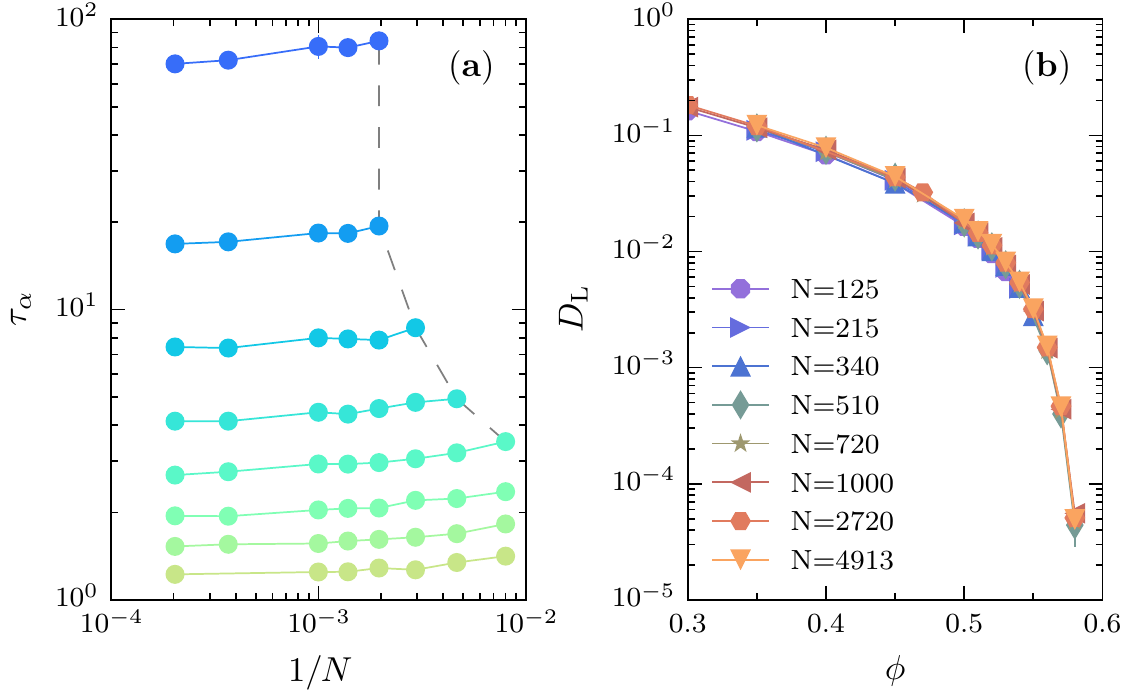}
  \caption{(a)~Dependence of the structural relaxation time $\ta$ on system size. Shown are the same packing fractions as in Fig.~\ref{fig:glass} with corresponding colors, from $\phi=0.50$ (bottom) to $\phi=0.57$ (top). Simulations carried out beyond the dashed line showed crystallization. (b)~In contrast, the long-time diffusion coefficient $\Dl$ is nearly system-size independent.}
  \label{fig:size}
\end{figure}

In Fig.~\ref{fig:size}(a) we plot the structural relaxation times $\ta$ as a function of system size (particles number $N$ at fixed packing fraction $\phi$). In agreement with similar results for molecular glass formers~\cite{karm09,bert12} and general arguments~\cite{bert12}, we observe that the relaxation time is approximately independent of $N$ for large systems and increases for small systems. The increase of $\ta$ for small $N$ at large $\phi$ is $\sim20$\% at most. In contrast, the long-time diffusion coefficient $\Dl(\phi)$ plotted in Fig.~\ref{fig:size}(b) shows virtually no dependence on system size. 

\subsection{Bond-order parameter}

Following Ref.~\citenum{stei83}, we employ a general bond-order parameter that is able to pick up structural order. To this end, to every particle with index $k$ the complex vectors $\mathbf q_l$ with components
\begin{equation}
  q_l^m(k) = \frac{1}{\Nb(k)}\sum_{k'=1}^{\Nb(k)} Y_l^m(\theta_{kk'},\vhi_{kk'})
\end{equation}
are assigned, where $Y_l^m(\theta,\vhi)$ are the spherical harmonics and the angles $\theta$ and $\vhi$ describe the orientation of the displacement vector between particles $k$ and $k'$ with respect to a fixed reference frame. Here, $\Nb(k)$ is the number of neighbors of particle $k$, where two particles are designated neighbors if their distance is smaller than 1.4 roughly corresponding to the first coordination shell. Setting $l=6$ for six-fold symmetries, we define the normalized scalar product
\begin{equation}
  S(k,k') = \frac{\mathbf q_6(k)\cdot\mathbf q_6^\ast(k')}{|\mathbf q_6(k)||\mathbf q_6(k')|}
\end{equation}
between two particles, where the asterisk denotes the conjugate complex and $0\leqslant S(k,k')\leqslant 1$. It can be interpreted as a bond strength between particles. The order parameter
\begin{equation}
  \hat\psi_6(k) = \frac{1}{\Nb(k)}\sum_{k'=1}^{\Nb(k)}S(k,k')
\end{equation}
then quantifies how strongly particle $k$ is bonded with its neighbors, assuming a value of unity in a perfect crystal (with six-fold symmetry) and a broad distribution around 0.2-0.3 in a disordered environment. Finally, we calculate the average value
\begin{equation}
  \psi_6(x) = \frac{1}{N}\sum_{k=1}^N \hat\psi_6(k)
\end{equation}
of all particles in a configuration $x=\{\x_k\}$ of particle positions.


\section{Kinetics of crystallization}

The freezing packing fraction for $s\simeq0.08$ estimated from Ref.~\citenum{faso03} is $\phi_\text{f}\simeq0.525$ (although for a different size distribution). For $N=340$ we study 12 different densities from $\phi=0.545$ to $\phi=0.578$, and for each of them we observed crystallization. For $N=510$ we study 3 densities (0.555, 0.570, 0.580): no crystallization was observed for $0.555$ within the simulation time (1 million time units). For packing fractions $\phi=0.570$ and $\phi=0.580$ crystallization occurred but at a much lower rate compared to $N=340$.

\begin{figure}[t]
  \centering
  \includegraphics{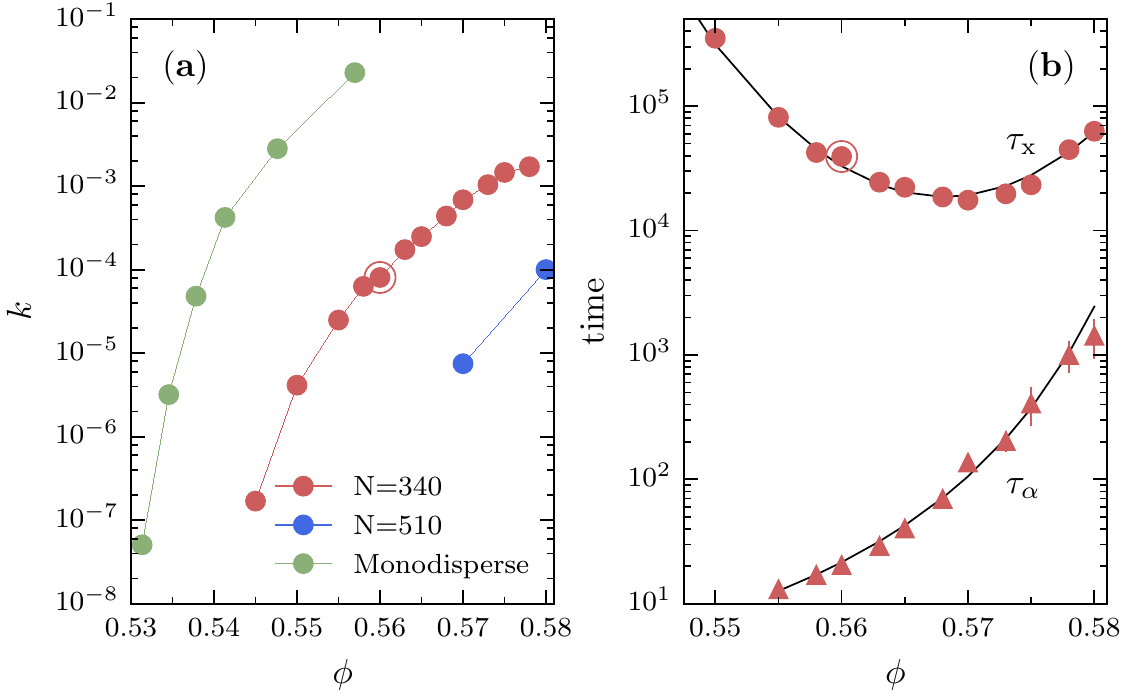}
  \caption{(a)~Crystallization rate density $k$ [Eq.~(\ref{eq:k})] normalized by the long-time diffusion coefficient $\Dl$ for two system sizes $N=340$ and $N=510$. For comparison, also shown is the data for monodisperse hard spheres from Filion \emph{et al.}~\cite{fili10}. (b)~Crystallization times $\tx$ and structural relaxation times $\ta$ as a function of packing fraction for $N=340$, cf. Fig.~\ref{fig:sketch}. Lines are guides to the eye, top line is a quadratic fit and bottom line is a fit to Eq.~(\ref{eq:tau}) with $\delta=1$. The highlighted points show the packing fraction that is studied in more detail in Sec.~\ref{sec:role}.}
  \label{fig:rates}
\end{figure}

Initial configurations are generated by compression starting from the equilibrated fluid at $\phi=0.3$. The packing fraction is increased in steps of 0.005 every $\Delta t=1$. For each packing fraction, we run 500 independent simulations and monitor the value of $\psi_6$. We stop a simulation run when $\psi_6=0.5$ is reached. The fraction $f(t)$ of surviving runs that at time $t$ have not yet crystallized is well described by the exponential decay $f(t)=e^{-t/\tx}$, from which we extract the crystallization time $\tx$. The normalized rate density is then
\begin{equation}
  \label{eq:k}
  k = \frac{1}{V\Dl\tx}.
\end{equation}
The rates as a function of packing fraction are plotted in Fig.~\ref{fig:rates}(a), which show a strong dependence on $N$. Also shown is the data for monodisperse hard spheres taken from Ref.~\citenum{fili10}. Even if shifted by the freezing packing fraction $\phi_\text{f}$, the rates will not collapse onto a common master curve.

\begin{figure}[t]
  \centering
  \includegraphics{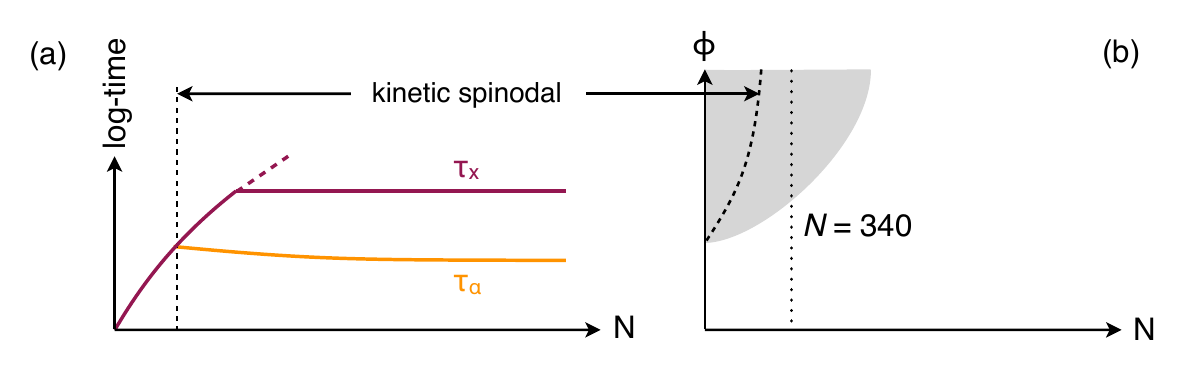}
  \caption{(a)~Sketch of the dependence of crystallization time $\tx$ and structural relaxation time $\ta$ on system size. While the crystallization time in sufficiently large systems should be nearly independent of $N$, there will be a crossover to a different behavior in small systems with $\tx$ decreasing. The structural relaxation time $\ta$ increases only modestly. The point where it crosses $\tx$ defines the kinetic spinodal, beyond which crystallization intervenes before the melt has relaxed. (b)~Sketch of the region (shaded area) within which crystallization is observable on simulation timescales, cf. Fig.~\ref{fig:size}(a). The dotted line indicates the system size studied in detail.}
  \label{fig:times}
\end{figure}

The crystallization times $\tx$ for $N=340$ are plotted in Fig.~\ref{fig:rates}(b) together with the structural relaxation times $\ta$. To calculate the latter, only runs that did not crystallize have been included to calculate the self-intermediate scattering function. While $\ta$ increases monotonously, $\tx$ shows a non-monotonous behavior in qualitative agreement with the sketch in Fig.~\ref{fig:sketch}. The minimum of $\tx$ at $\phi\simeq0.57$ can be interpreted as the crossover between two qualitatively different, collective relaxation mechanisms to reach the stable solid. For smaller densities $\phi<0.57$ crystallization proceeds by nucleation and growth. There is an entropic cost $\Delta F^\ddagger$ independent of $N$ for forming stable solid clusters, which has to be overcome by thermal fluctuations. Classical nucleation theory predicts a form $\tx\sim(1/V)e^{\Delta F^\ddagger}$ for the crystallization time. If the system is large enough multiple independent nucleation events occur, which is accounted for by the pre-factor, leading to a weak logarithmic dependence of $\ln\tx$ on particle number $N$. For large densities, small solid domains form but coarsening is inhibited. Also in this regimes $\ln\tx$ only weakly depends on $N$. Going to small systems, there will be a crossover to a different behavior since thermal fluctuations can probe a larger fraction of the configuration space and thus allow access to crystalline configurations on times $\ln\tx\sim\Delta F\sim N^{2/3}$ with free energy barrier $\Delta F$ associated with an interface, see Fig.~\ref{fig:times}. All our numerical results qualitatively agree with this simple picture. The system size $N=340$ crosses the region where crystallization is observable but the separation of $\ta$ and $\tx$ is still large so that the concept of a metastable fluid is meaningful. While the scaling of $\tx$ depends on $N$, the transition configurations comprising the top of the barrier should be qualitatively similar to those in larger systems.

\section{Role of local structure}
\label{sec:role}

\subsection{Committor analysis}

\begin{figure}[t]
  \centering
  \includegraphics{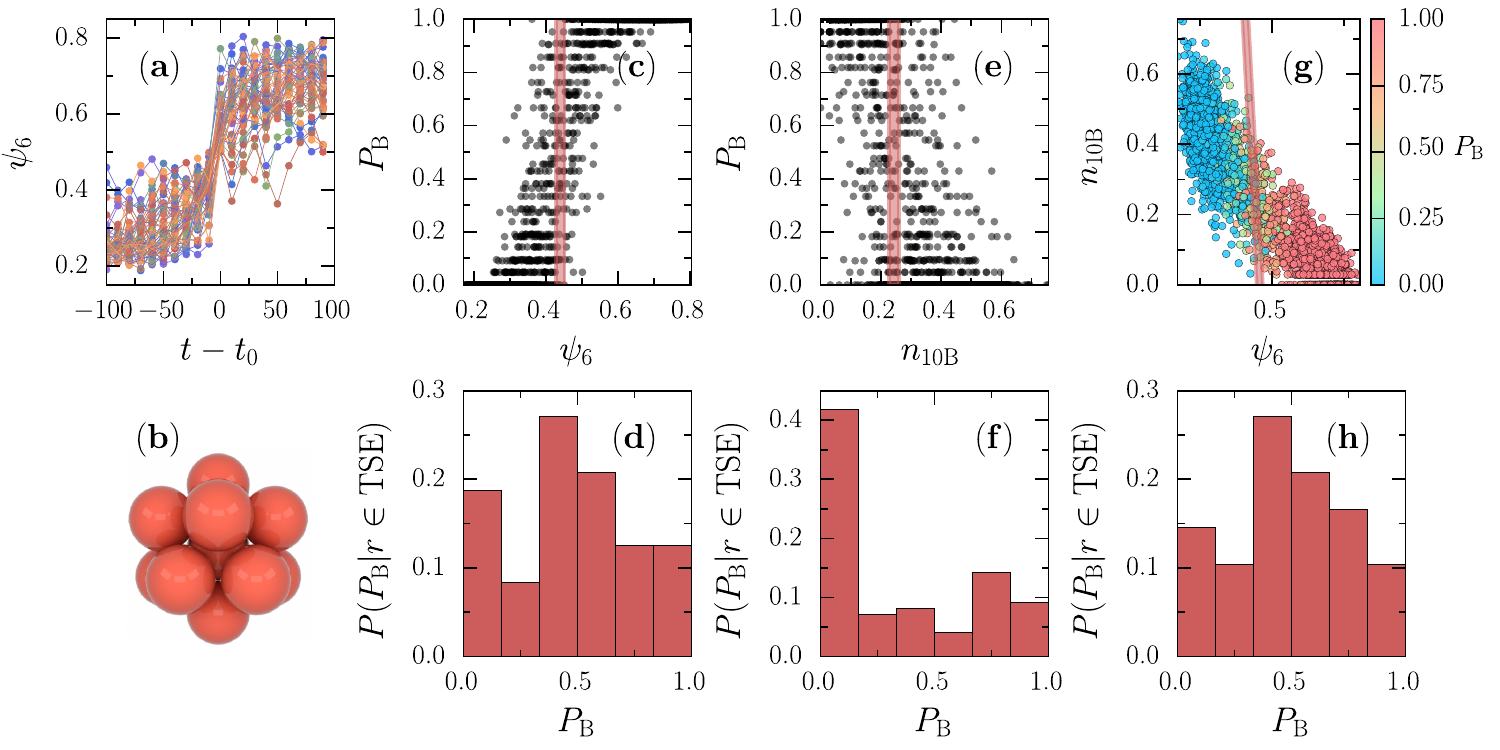}
  \caption{Committor analysis for $\phi=0.560$. (a)~Time series of the bond-order parameter $\psi_6$ for reactive trajectories shifted by the time $t_0$ when crossing $\psi_6=\frac{1}{2}$. (b)~Snapshot of a 10B cluster. (c)~Scatter plot of bond order $\psi_6$ \emph{vs.} commitment probability $\pb$. Every symbol corresponds to one configuration. The shaded area indicates the configurations of the transition state ensemble (TSE). (d)~Histogram of $\pb$ for the TSE configurations. (e,f)~Same as (c,d) but for the population $n_\text{10B}$ of 10B clusters. (g)~Scatter plot of bond order $\psi_6$ \emph{vs.} the population $n_\text{10B}$, where colors correspond to $\pb$. Indicated is again the TSE obtained from the likelihood maximization with (h)~histogram of $\pb$.}
  \label{fig:commit}
\end{figure}

We perform a committor analysis~\cite{dell02} to access the transition configurations and to gain insight into the microscopic pathway that the melt takes to crystallize. We do this for the packing fraction $\phi=0.560$ with $N=340$ particles, for which crystallization is an activated process -- \emph{i.e.}, a sudden transition occurs after a waiting time -- but for which the average waiting time is not too long. To this end, a set of reactive trajectories is collected, selecting a time window $\Delta t$ around the nucleation event occurring at $t_0$ defined through $\psi_6(t_0)=0.5$, see Fig.~\ref{fig:commit}(a). For each configuration $x_l=x(t_l)$ stored at times $t_l$ along the trajectory $x(t)$, a number $N_l$ of ``fleeting'' trajectories of length $\tf$ is generated by changing randomly the velocities of the particles (distributed according to a Maxwellian). The fleeting time $\tf=10$ is chosen as the average time it takes for the system to complete the transition starting from the fluid state.

A fleeting trajectory may or may not ``commit'' to the solid state: for each configuration $x_l$ we estimate the commitment probability $\pb(x_l)$ through the ratio $\frac{\#\text{commits}}{N_l}$. The commitment probability constitutes the exact reaction coordinate for the transition. The ensemble of configurations for which $\pb\simeq\frac{1}{2}$ is termed the \emph{transition state ensemble} (TSE). Typically, the full function $\pb(x)$ is too expensive to be computed explicitly and, moreover, it does not give insights into the microscopic details of the transition. Hence, it is often preferable to find an approximate reaction coordinate $r(\{q_i\})$ involving a combination of collective variables $q_i(x)$ as order parameters. Without loss of generality, we define $r_\ast=0$ to correspond to the TSE. Close to the transition, we approximate the reaction coordinate by the linear combination
\begin{equation}
  r(x) = \al_0 + \sum_i \al_i q_i(x)
\end{equation}
with unknown coefficients $\{\al_i\}$.

\subsection{Order parameters}

To study the crystallization of hard spheres, we employ the bond-order parameter $q_i=\psi_6$ and the populations $q_i=n_i$ of various local structures. Note that due to the small system size we do not attempt to identify the largest cluster of solid particles. The local motifs are detected employing the topological cluster classification method (for technical details see Ref.~\citenum{mali13}), which is based on a modified Voronoi tessellation and the detection of $n$-membered rings of particles. Specifically, we look for a fivefold symmetric arrangement of 10 particles termed ``10B'' (the nomenclature follows Ref.~\citenum{doye95} for minimum energy clusters of the Morse potential). The population $n_\text{10B}(x)=N_\text{10B}/N$ is the fraction of particles (independent of their diameter) participating in this structural motif. It has been found in Ref.~\citenum{taff13} that the population of this motif increases as the packing fraction is increased. It is abundant in the metastable fluid but the population strongly drops upon crystallization. We also consider 9X, which is highly populated for FCC and BCC crystal structures (but also occurs in the fluid), and 7A (a pentagonal bipyramid), which should only be populated in the fluid.

\subsection{Likelihood maximization}

We now aim to determine which linear combination of order parameters fits best the observed reactive trajectories. To this end, we use a maximum likelihood approach~\cite{pete06}. Specifically, we follow the approach of Ref.~\citenum{jung13}.

For one configuration $x_l$, the probability to observe $m_l$ (out of $N_l$) fleeting trajectories that commit is given by a binomial distribution. The probability for a single trial is $\pb(x_l)\approx\pbm(r_l)$ with $r_l=r(x_l)$, which we model through
\begin{equation}
  \pbm(r) = \frac{1}{2}[1+\tanh(r)].
\end{equation}
This is a generic function that smoothly interpolates between negative values of the reaction coordinate $r$ for which transitions are unlikely, $\pb\simeq0$, and positive values for which $\pb\simeq1$. Assuming $\Nc$ independent configurations then leads to the probability
\begin{equation}
  P(\{m_l\}|\{\al_i\}) = \prod_{l=1}^{\Nc} {N_l \choose m_l} [\pbm(r_l)]^{m_l}[1-\pbm(r_l)]^{N_l-m_l}.
\end{equation}
Given the sequence $\{m_l\}$ obtained from the simulated fleeting trajectories, the likelihood function $L(\{\al_i\})=P(\{m_l\}|\{\al_i\})$ is a function of the expansion coefficients $\{\al_i\}$. It describes the suitability of a model defined through $\{\al_i\}$ given the observed data $\{m_l\}$. Practically, one considers the log-likelihood and defines as cost function
\begin{equation}
  \label{eq:cost}
  C(\{\al_i\}) = \ln L(\{\al_i\}) - \frac{n_\al}{2}\ln \Nc,
\end{equation}
where $n_\al$ is the number of parameters $\al_i$. More complex models involving more order parameters have higher values for the likelihood, which is compensated by subtracting the second term to make models with different values for $n_\al$ comparable ~\cite{schw78}. The cost function~(\ref{eq:cost}) is maximized using the Nelder-Mead algorithm, yielding the coefficients $\{\al_i\}$.

\subsection{Results}

\begin{table}[t]
  \centering
  \begin{tabular}{c|c|c|c|c|l}
    $\psi_6$ & $n_{10\mathrm{B}}$ & $n_{9\mathrm{X}}$ & $n_{7\mathrm{A}}$ & $n_\al$ & $C(\psi_6)/C$ \\
    \hline
    $\bullet$ & $\bullet$ & & &                           2 & 1.0026(2) \\
    $\bullet$ & $\bullet$ & $\bullet$ & &                 3 & 1.0021(2) \\
    $\bullet$ & & & &                                     1 & 1.0000(2) \\ 
    $\bullet$ & & & $\bullet$ &                           2 & 0.9995(2) \\ 
    $\bullet$ & & $\bullet$ & &                           2 & 0.9993(2) \\ 
    $\bullet$ & & $\bullet$ & $\bullet$ &                 3 & 0.998(7)  \\ 
    $\bullet$ & $\bullet$ & & $\bullet$ &                 3 & 0.99(2)   \\ 
    $\bullet$ & $\bullet$ & $\bullet$ & $\bullet$ &       4 & 0.98(2)   \\ 
    \hline
             & $\bullet$ & $\bullet$ & $\bullet$ &        3 & 0.879(4)  \\ 
             & & $\bullet$ & $\bullet$ &                  2 & 0.8661(5) \\ 
             & $\bullet$ & & $\bullet$ &                  2 & 0.8480(1) \\ 
             & & &$\bullet$ &                             1 & 0.8418(1) \\ 
             & $\bullet$ & $\bullet$ & &                  2 & 0.7293(1) \\ 
             & & $\bullet$ & &                            1 & 0.6750(1) \\ 
             & $\bullet$ & & &                            1 & 0.6154(1) \\ 
  \end{tabular}
  \caption{Results for the likelihood maximization computed for several models, each involving a different combination of order parameters as indicated by the black dots. Models are sorted by their $C$ values normalized by the value for the $\psi_6$ model. In brackets we report uncertainties, which are obtained by bootstrapping the data. The bond-order parameter $\psi_6$ outperforms local structures as reaction coordinate, with the best model being the combination of $\psi_6$ and $n_\text{10B}$.}
  \label{tab:results}
\end{table}

Fig.~\ref{fig:commit}(c-f) show the results of the likelihood maximization using one order parameter ($n_\al=1$): (c,d) for $\psi_6$ and (e,f) for $n_\text{10B}$. The transition state value of the order parameter is then given by $-\al_0/\al_1$. As TSE we collect all configurations with $-0.15<r<+0.15$. In Fig.~\ref{fig:commit}(d,f) we show the corresponding histograms $P(\pb|r\in\text{TSE})$ of committor probabilities. For a good reaction coordinate $r$ one expects these distributions to be symmetric and peaked around $\pb=\frac{1}{2}$~\cite{humm04}. While this is approximately the case for $\psi_6$, the histogram for $n_\text{10B}$ is rather flat with a peak for small values of $\pb$. Hence, $n_\text{10B}$ performs quite poorly as a reaction coordinate. This is also reflected in the values $C$ of the cost function provided in Table~\ref{tab:results}, which lists all models studied.

We systematically tested different combinations of order parameters with results given in Table~\ref{tab:results}, which are ranked by their values for $C$. To estimate the uncertainty of $C$, we have bootstrapped the data. The bootstrapping is performed by resampling the values for $\pb$ adding normal noise and repeating the maximization procedure. The variance of the Gaussian noise is set equal to the residual between the data and the model obtained by maximization with zero noise. Note that all combinations of local structures 10B and 7A have large uncertainties, presumably because the joint distribution is not unimodal.

In Table~\ref{tab:results} we observe a separation between the performance of local structures alone as reaction coordinate, and the performance of models including $\psi_6$, which all have larger values of $C$. This means that bond-order better captures the transition states separating fluid from solid. Moreover, combining bond-order with local structure does not improve the reaction coordinate, from which we conclude that these local structures add little information to the description of these transition states. The only exception is 10B, which slightly improves $\psi_6$ and turns out to be the best model. The results for this combination are also shown in Fig.~\ref{fig:commit}(g,h). In the plane spanned by both order parameters, the TSE is now a line, with histogram of committor probabilities shown in Fig.~\ref{fig:commit}(h).


\section{Conclusions}

For hard spheres with polydispersity of about 8\% we find that the structural relaxation time $\ta$ of the supersaturated melt increases for small systems composed of a few hundred particles, but that this increase is moderate with at most $\sim10$\% for $N=340$. Single particle motion, as captured by the long-time diffusion coefficient, is independent of system size for the range of sizes and packing fractions studied here. In contrast, the crystallization rate density $k$ strongly depends on system size. For $N=340$ crystallization events can be studied in straightforward computer simulations but already for $N=510$ the rate is about two orders of magnitude smaller, making a systematic study of crystallization unfeasible. Hence, we conclude that hard spheres with $s\simeq0.08$ crystallize but that the kinetics slows down dramatically compared to smaller $s$.

Focusing on a small system with $N=340$ particles, we find the expected qualitative behavior for crystallization time $\tx$ and structural relaxation time $\ta$, cf. Fig.~\ref{fig:sketch} and Fig.~\ref{fig:rates}(b). In particular, while $\ta$ increases monotonously, $\tx$ first decreases and then again increases. We interpret the smallest crystallization time around $\phi\simeq0.57$ to indicate the crossover from activated nucleation and growth at smaller packing fractions to a regime in which crystallization kinetics is limited by diffusion. Finally, we have performed a committor analysis for $\phi=0.560$, which lies in the activated regime but is close to the crossover. For this packing fraction we expect a small critical nucleus and in small systems collective fluctuation that reach this barrier are more likely. We find that global bond-order is a good reaction coordinate and that most local structures fail to capture the transition from fluid to crystalline. The best reaction coordinate is found to be the combination of the bond-order parameter and the population of 10B-clustered particles. While the conversion of local structures during the crystallization process is insightful~\cite{taff13}, it remains to be seen why they do not seem to matter for the transition state.

To conclude, we argue that small systems allow computational insights into the same mechanisms that are at play in large systems. This appears to be a fruitful but often neglected avenue to systematically study model glassformers. One route to extend the range of system sizes considered here is to employ rare-event techniques like forward flux sampling~\cite{alle09}. Another interesting question is how hard-sphere glasses crystallize under shear~\cite{blaa04,land13,rich15}.


\ack

M.C. is funded by the DFG through the Graduate School ``Materials Science in Mainz'' (GSC 266) and the collaborative research center TRR 146. We thank David Richard for discussions and help with the committor analysis.


\providecommand{\newblock}{}

\end{document}